# Remarks on the Order Parameter of the Cuprate High Temperature Superconductors

Stefan Hüfner [*)], Frank Müller

*Experimental Physics, Saarland University, Campus C6.3, 66123 Saarbrücken, Germany*

A large body of spectroscopic data on the cuprate high temperature superconductors (CHTSC) is reviewed in order to determine their order parameter. ASJ, INS, $B_{2g}$ Raman spectra, optical data, NIS „dips", ARPES „dips" and ARPES „kinks" all show the same excitation energy (40 meV for OP95 systems), proportional to the superconducting transition temperature, and it is therefore identified with the order parameter.

[*)] *Corresponding author:* huefner@mx.uni-saarland.de

## Introduction

Even 20 years after their discovery [1], the Cuprate High Temperature Superconductors (CHTSC) are not understood. A central property that characterizes a conventional superconductor is the gap that opens below $T_c$. [2,3] It measures the binding energy of the Cooper pairs and is equal to the condensation energy which is the order parameter.

In the CHTSC, the situation is different. [4-12] In these systems there are two gaps. One gap extends above $T_c$ up to $T^*$ (pseudogap, pg). It is difficult to identify it with the order parameter because it exhibits a linear rather than a parabolic doping dependence. A parabolic doping dependence is established for $T_c$ which in turn should be proportional to the order parameter, meaning that a parabolic doping dependence is to be expected for the order parameter. In addition, there is the superconducting gap (scg), and it seems not clear whether it evolves smoothly out of the pg or whether it is a gap in itself different from the pg. [5-7, 10, 11] In the literature, the connection of the two gaps is discussed controversially, where one line of thought identifies the pg as the precursor to the scg (preformed pairs) [13], whereas another line of argument identifies the pg as the manifestation of a competing order. [14]

While the pg is well established through different types of experiments [4] like ARPES [10], NIS [15] or STM tunnelling [11], this does not hold for the scg. In going through the literature, one finds only one type of experiment that states explicitly, that it measures the scg in the CHTSC, namely Andreev Saint James reflection (ASJ). [5] INS [16] and optical spectroscopy [17] measure a magnetic resonance, $B_{2g}$ Raman spectroscopy [12] measures a nodal gap and the sidebands („dips") as observed in NIS [15] and ARPES [10] as well as the „„kinks" [10] in the latter technique are called just that.

In some cases, the gap measured as the pg above $T_c$ is termed the scg below $T_c$. [10, 11] This seems a questionable procedure, because, as mentioned above, the pg has a linear doping dependence while $T_c$ has a parabolic doping dependence. Since the order parameter (or for the present purpose: the scg) should follow $T_c$, it is unlikely that the pg is the scg below $T_c$.

In this controversial situation, it seems to be useful to compare a large body of experimental data in order to find general trends with respect to the scg. It will be shown that there is strong evidence, that the 40 meV excitation in 95 K OP CHTSC (and related excitations in other CHTSC) which has been termed the scg, a „dip", a „kink", an optical resonance or a magnetic resonance in INS, is always the same excitation, namely the order parameter. It measures the same excitation in all these different experiments, namely the energy needed to excite a pair out of the condensate (or the energy gained in putting it into the condensate).

## Analysis of experimental data

In Fig. 1, the optical data of the magnetic resonance for a number of CHTC, as given by Fig. 4 of Ref. 18, are compared to the INS data for the magnetic resonance in Y123 [16], $B_{2g}$ Raman data for Hg 1201 [19], ASJ data for Y123 [5], NIS „dip" data for Bi2212 [15], ARPES „dip" data for Bi2201 [20, 21] and data for the antinodal „kink" in Bi2212. [10, 22-24] While there is scatter in the data, it seems fair to state, that they follow a common trend. Thus, it is not unreasonable to assume that this excitation has a common origin. Since ASJ data [5] can be interpreted as measuring the scg, all the other techniques should measure the same quantity.

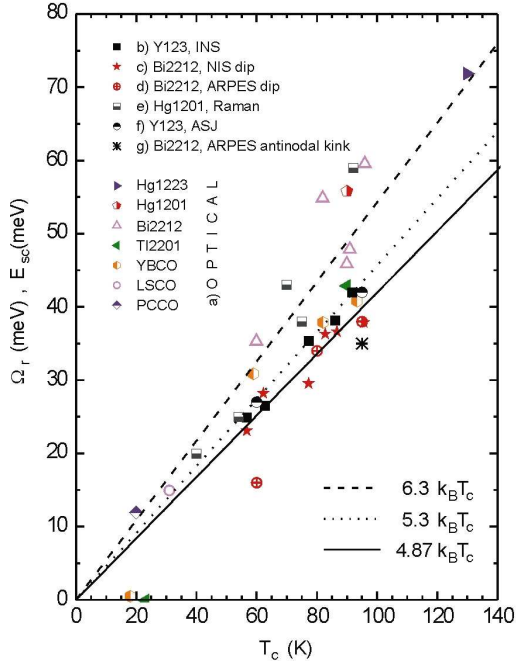

**Figure 1 (color online):** Order parameter of CHTCs obtained by different measuring techniques. a) optical data according to Ref. 18 (and references therein), b) INS on Y123 Ref. 16, c) NIS on Bi2212 Ref. 15, d) ARPES „dip" on Bi2212 Ref. 20, e) Raman $B_{2g}$ on Hg1201 Ref. 19, f) ASJ on Y123 Ref. 5, g) ARPES antinodal „kink" on Bi2212 Ref. 22.

We are fully aware of the fact that this designation does not agree with much of the literature and therefore, we want to discuss the data for the order parameter as obtained by different techniques.

The Raman data have been very transparently analysed by LeTacon et al. [19] They argue, that the nodal $B_{2g}$ and the antinodal $B_{1g}$ Raman response derives from the same matrix element and that only the direction specific properties are different for the two orientation. However, the authors themselves point out, that in this context the different temperature dependence of the nodal and the antinodal signal cannot be explained. In a later study [25] the authors address the constant ratio of $T_c$ and the $B_{2g}$ signal and they suggest, that the $B_{2g}$ signal may be the order parameter

In this situation, the only solution seems to be the assumption, put forward by the present authors [6,7] that the antinodal Raman signal is indeed, as assumed by present theory [12], a pair breaking process measuring the antinodal pg, while the so-called nodal gap is a different type of transition, measuring the energy to take a boson out of the condensate and thus measuring the order parameter. A similar situation holds for the neutron resonance. [16] It is generally taken as a spin resonance where the exact nature of this resonance is under debate. There is however a study [26] from which it could be concluded, that the neutron resonance is the order parameter. A strong support for this interpretation comes from the fact, that the resonance peak looses much of its intensity in approaching $T_c$ from below. The interpretation of the peak in the real part of the optical conductivity as originating from the excitation of a boson out of the condensate poses no problem. Again, the energy and the temperature dependence support this assignment. [18]

With respect to the interpretation of the „dips" in NIS and ARPES spectra and the „kinks" measured in ARPES dispersion an interpretation is in principle straightforward.

The interpretation of the "dips" and "kinks" can follow the analysis given for the phonon sidebands observed in the tunnelling spectra in the conventional superconductors. [3, 27, 28] There, it was shown that the electron phonon interaction leads to sidebands in the tunnel spectra, from which in turn the phonon density of states can be recovered.

Looking at this phenomenon in a more general sense, one realizes that an elementary excitation in a solid like a plasmon, a phonon or a magnon can interact with the electrons. This leads to a change (renormalization) in the electron dispersion. From this renormalization the spectrum of the elementary excitation can be determined.

In a photoelectron spectrum there are two modes that allow the determination of the coupling to the elementary excitations. For a particular point in the Brillouin zone, one can either measure the spectrum of the "main line" and the adjacent satellite, which represents the signature of the elementary excitation. In the tunnel data and the ARPES spectra this feature is generally called a „dip". This is the common way in which the phonon sidebands are determined in tunnel spectra of conventional superconductors.

In another measuring mode the dispersion relation of a band is measured and the deviation from the unrenormalized band allows the extraction of the spectrum of the elementary excitation. In the ARPES data this is called the „kink". This means, e.g., that in the antinodal region the „dip" and the „kink" in the ARPES spectra have the same origin because they measure the same excitation in two different modes. [10]

In order to explain the nature of the „dips" observed in the tunnelling and ARPES data and the "kinks" measured in the ARPES dispersion of the CHTSCs, one needs a picture of the excitations (beyond phonons), which can couple to the electrons and thus lead to a renormalization.

For this purpose a simple electronic structure of the CHTSC is constructed.

For $T>T^*$ the CHTSC are Fermi liquids.

For the temperature regime $T^*>T>T_c$, a second phase develops, namely the pg liquid, where part of the Fermions couple to form pairs. This is reflected by the pg which in turn can couple to the electrons. Finally, for $T_c>T$, the pg pairs can condense into the superconducting condensate and the condensation energy is another elementary excitation. In the model used here, namely that of the preformed pairs, the pg liquid couples directly to the condensate.

ARPES in the Fermi liquid regime leads to the band structure.

ARPES in the second regime (pg liquid) leads to the band structure, modified by the pg. In this regime, the Fermi liquid signature (density of states at $E_F$) is only seen around the nodal point. The electrons photoemitted from the electronic states near the Fermi energy can be renormalized by the coupling to the pg liquid. This leads to the 70 meV „kink", seen in the nodal band structure of OP95 CHTSCs. [29] Note that the pseudogap energy of these systems is 70 meV. [6]

In the superconducting state, the ARPES (or the NIS) process at the antinodal point in a OP95 CHTSC breaks up a boson, leading to the pg gap energy of 35 meV (energy of the electron left behind). In addition, the process lifts the superconducting boson out of the condensate, which in these system needs 40 meV, which is the energy measured directly by ASJ [5], $B_{2g}$ Raman [12] NIS [15], and optical conductivity. [18] This leads to a "dip" energy of 40 meV for OP95 systems and to a correspondingly smaller one for UD or OD samples. [20]

Note that the „kink" at the antinodal point has about the same energy as the „kink" at the nodal point [10, 22, 24], but it shows a temperature dependence with a large reduction around $T_c$. [24] This reflects the coupling to the condensation energy of 40 meV which disappears at $T_c$. In this case, one has to take into account for the energy balance the pg gap energy, which leads to a partitioning of the „kink" energy in the following form: 75 meV = 35 meV+ 40 meV, The fact that the nodal and the antinodal „kink" have about the same energy is an accident, caused by the fact that in the systems under consideration one has $2\Delta_{pg} \sim 2E_r$ (see Fig.2 in Ref. 6) at optimal doping.

The data for „dips" and „kinks" vary considerably. This can be attributed to the fact that these spectral features have energies only of the order of 100 meV and that the bare spectra that have to be subtracted from the measured spectra in order to obtain the interaction energies are difficult to obtain.

## Summary and conclusions

The present study shows that the spectroscopic results for the CHTSC can be reconciled with each other in the framework of the preformed pairs model [5, 6, 13, 32-39] where there is now strong evidence for the existence of these preformed pairs. [30, 31] The pg phase consists of the Fermi liquid and the preformed pairs liquid, and below $T_c$, the pairs condense into the superconducting phase. This picture gives rise to two distinctly different excitations. A pair breaking excitation which measures the binding energy of the pairing interaction, above and below $T_c$, with a linear doping dependence, represented by the pg. Because of its linear doping dependence which does not follow the superconducting transition temperature it cannot be the order parameter. The second excitation, $E_{sc} = 2\Delta_{sc}$, occurs below $T_c$ and measures the binding energy of the pairs in the condensate.

This point is emphasized with the collection of data in Fig. 1. [18] In this diagram, the superconducting energy (twice the gap energy), as measured by various techniques, is plotted as a function of $T_c$ and the observed linear relation shows that the superconducting energy is the order parameter. [5, 6]

## Acknowledgement

This work was initiated during a stay of SH at the University of British Columbia in Vancouver, where stimulating discussions with A Damascelli, W. Hardy and G. Sawatzky took place. Important communications with C. Gross, J. Fink and B. Keimer are gratefully acknowledged.